\documentclass[reprint,superscriptaddress,aps,prl]{revtex4-1}%
\usepackage{graphicx}
\usepackage{dcolumn}
\usepackage{bm}
\usepackage{bbold}
\usepackage{amsfonts,color}
\usepackage{amsmath}
\usepackage{verbatim}%
\usepackage{amsfonts}%
\usepackage{amssymb}
\providecommand{\U}[1]{\protect\rule{.1in}{.1in}}

\newcounter{lastnote}

\begin{document}

\title{Nonlocal detection of interlayer three-magnon coupling}
\date{\today}
\author{Lutong Sheng}
\thanks{These authors contributed equally to this work.}
\affiliation{Fert Beijing Institute, MIIT Key Laboratory of Spintronics, School of Integrated Circuit Science and Engineering, Beihang University, Beijing 100191, China}
\author{Mehrdad Elyasi}
\thanks{These authors contributed equally to this work.}
\affiliation{WPI Advanced Institute for Materials Research, Tohoku University, Sendai 980-8577, Japan}
\author{Jilei Chen}
\thanks{These authors contributed equally to this work.}
\affiliation{Shenzhen Institute for Quantum Science and Engineering, Southern University of Science and Technology, Shenzhen 518055, China}
\affiliation{International Quantum Academy, Shenzhen 518048, China}
\author{Wenqing He}
\thanks{These authors contributed equally to this work.}
\affiliation{Beijing National Laboratory for Condensed Matter Physics, Institute of Physics, University of Chinese Academy of Sciences, Chinese Academy of Sciences, Beijing 100190, China}
\author{Yizhan Wang}
\affiliation{Beijing National Laboratory for Condensed Matter Physics, Institute of Physics, University of Chinese Academy of Sciences, Chinese Academy of Sciences, Beijing 100190, China}
\author{Hanchen~Wang}
\affiliation{Fert Beijing Institute, MIIT Key Laboratory of Spintronics, School of Integrated Circuit Science and Engineering, Beihang University, Beijing 100191, China}
\affiliation{International Quantum Academy, Shenzhen 518048, China}
\author{Hongmei Feng}
\affiliation{State Key Laboratory of Low-Dimensional Quantum Physics and Department of Physics, Tsinghua University, Beijing 100084, China}
\author{Yu Zhang}
\affiliation{Beijing National Laboratory for Condensed Matter Physics, Institute of Physics, University of Chinese Academy of Sciences, Chinese Academy of Sciences, Beijing 100190, China}
\author{Israa Medlej}
\affiliation{Shenzhen Institute for Quantum Science and Engineering, Southern University of Science and Technology, Shenzhen 518055, China}
\affiliation{International Quantum Academy, Shenzhen 518048, China}
\author{Song Liu}
\affiliation{Shenzhen Institute for Quantum Science and Engineering, Southern University of Science and Technology, Shenzhen 518055, China}
\affiliation{International Quantum Academy, Shenzhen 518048, China}
\author{Wanjun Jiang}
\affiliation{State Key Laboratory of Low-Dimensional Quantum Physics and Department of Physics, Tsinghua University, Beijing 100084, China}
\author{Xiufeng Han}
\affiliation{Beijing National Laboratory for Condensed Matter Physics, Institute of Physics, University of Chinese Academy of Sciences, Chinese Academy of Sciences, Beijing 100190, China}
\author{Dapeng Yu}
\affiliation{Shenzhen Institute for Quantum Science and Engineering, Southern University of Science and Technology, Shenzhen 518055, China}
\affiliation{International Quantum Academy, Shenzhen 518048, China}
\author{Jean-Philippe Ansermet}
\affiliation{Institute of Physics, Ecole Polytechnique F\'ed\'erale de Lausanne (EPFL), 1015, Lausanne, Switzerland}
\affiliation{Shenzhen Institute for Quantum Science and Engineering, Southern University of Science and Technology, Shenzhen 518055, China}
\author{Gerrit E. W. Bauer}
\affiliation{WPI Advanced Institute for Materials Research, Tohoku University, Sendai 980-8577, Japan}
\affiliation{Institute for Materials Research, Tohoku University, Sendai 980-8577, Japan}
\affiliation{Center for Spintronics Research Network, Tohoku University, Sendai 980-8577, Japan}
\affiliation{Zernike Institute for Advanced Materials, University of Groningen, 9747 AG Groningen, Netherlands}
\author{Haiming Yu}
\email{haiming.yu@buaa.edu.cn}
\affiliation{Fert Beijing Institute, MIIT Key Laboratory of Spintronics, School of Integrated Circuit Science and Engineering, Beihang University, Beijing 100191, China}
\affiliation{International Quantum Academy, Shenzhen 518048, China}

\begin{abstract}
A leading nonlinear effect in magnonics is the interaction that splits a high-frequency magnon into two low-frequency ones with conserved linear momentum. Here, we report experimental observation of nonlocal three-magnon scattering between spatially separated magnetic systems, \textit{viz}. a CoFeB nanowire and an yttrium iron garnet (YIG) thin film. Above a certain threshold power of an applied microwave field, a CoFeB Kittel magnon splits into a pair of counter-propagating YIG magnons that induce voltage signals in Pt electrodes on each side, in excellent agreement with model calculations based on the interlayer dipolar interaction. The excited YIG magnon pairs reside mainly in the first excited ($n=1$) perpdendicular standing spin-wave mode. With increasing power, the $n=1$ magnons successively scatter into nodeless ($n=0$) magnons through a four-magnon process. Our results help to assess non-local scattering processes in magnonic circuits that may enable quantum entanglement between distant magnons for quantum information applications.
\end{abstract}

\maketitle

Nonlinear effects are ubiquitous in a variety of physical systems, such as lasers~\cite{Haen2002}, electron beams~\cite{Mo2013}, cold atoms~\cite{Dut2001}, and water waves~\cite{Roz2022}. Magnons are the quanta of spin waves, the collective excitations of the magnetic order. In magnon spintronics or magnonics~\cite{Dirk2010,Chu2015,Vla2008,vanWees2015,Pirro2021} they are employed as information carriers for low-power processing and transmission~\cite{Khitun2010,Csaba2016,QWang2020}. Non-linearities in the magnetization dynamics are known for many decades~\cite{Suhl1957,Wigen_book,Rez1996,Demo2003,Patton2009,Demo2011,Ando2012,Buch2019,Helmut2009,Cam2014}. New nonlinear phenomena have been discovered in magnetic textures~\cite{KSchu2019,KSchu2020,Slavin2021,Yan2021,Wol2022}, nanoscale magnets~\cite{Ilya2019,Demi2019,Moh2021} and hybrid systems~\cite{You2018,Mehrdad2020}. Non-linearity provides rich physics~\cite{Kon2016} and is relevant for technological applications~\cite{Jae2004}, such as mechanical force sensors~\cite{Cra2000}. Optical non-linearities enable the coupling of LC resonators with a superconducting qubit~\cite{Adh2013}. A nonlinear magnetostrictive interaction may generate magnon-photon-phonon entanglement in cavity magnetomechanics~\cite{JLi2018}. Non-linearities generate magnon interaction that leads to \textquotedblleft squeezing\textquotedblright\ of magnon amplitudes~\cite{Yuan2020} and continuous variable quantum entanglement~\cite{Naka2020}. The latter is distillable~\cite{Ecker2021} only when non-local and quantum-entangled over a distance. Separation of entangled magnons appears to be a formidable task, but could be useful in information technologies such as quantum key distribution~\cite{Sca2009} and quantum teleportation~\cite{Pan2017}.

\begin{figure}[ht]
\centering
\includegraphics[width=86mm]{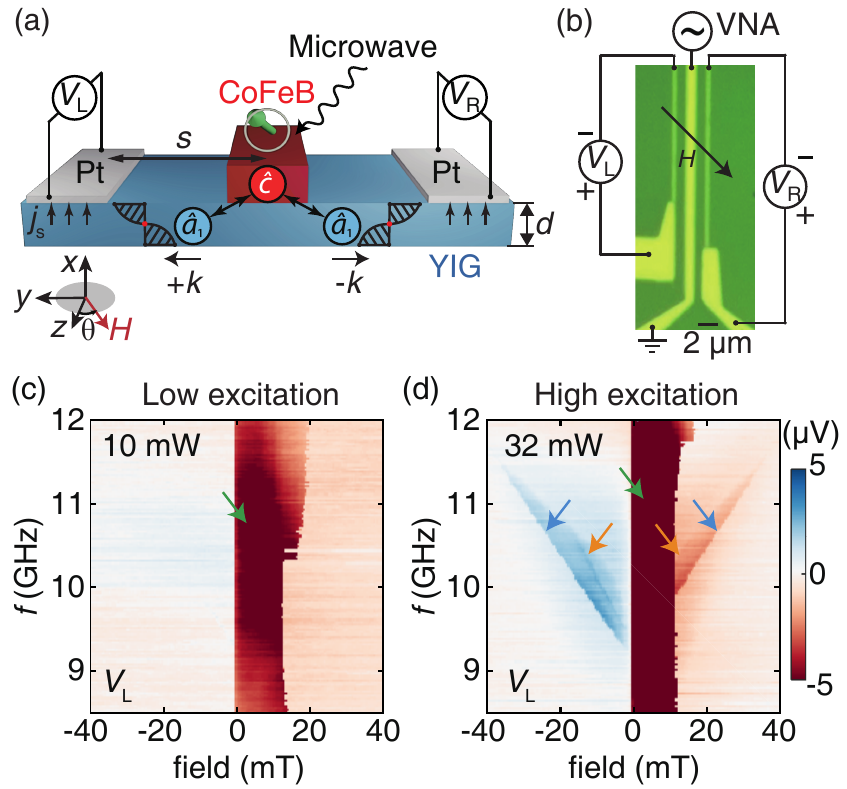}  
\caption{(a) Schematic (side view) of spin pumping by a CoFeB nanowire detected non-locally by two Pt electrodes on the left ($V_{\text{L}}$) and right ($V_{\text{R}}$) sides. Magnetic field $H$ is applied in the $yz$ plane with an angle $\theta=45^{\circ}$ with respect to the $z$ axis. $s$ denotes the distance between the CoFeB wire and a Pt bar. (b) Optical microscopic image of the three-terminal device. The center line is a CoFeB nanowire covered by a gold microwave stripline antenna. The Pt electrodes on both sides detect nonlocal spin pumping from the CoFeB wire. (c) $V_{\text{L}}$ measured with field swept from negative to positive values. At an input microwave power of 10~mW, we are still in the linear regime. The green arrow marks the spin pumping when the magnetizations are anti-parallel. (d) $V_{\text{L}}$ measured at a high input power of 32~mW. The field sweep is the same as in (c). Light blue arrows indicate a mode associated with the interlayer three-magnon coupling. We attribute another mode (orange arrows) to a secondary nonlinear process.}%
\label{fig1}%
\end{figure}
The leading nonlinear effect in magnetic systems is the three-magnon interaction~\cite{Patton2009,Demo2011,Ando2012,Buch2019,Helmut2009,Cam2014,KSchu2019,KSchu2020,Slavin2021}, in which a magnon with energy $\hbar\omega$ and zero momentum (Kittel mode) decays into two lower energy (frequency) magnons $\hbar\omega/2$ with opposite wave vectors ($\hbar k$ and $-\hbar k$). It has been observed in yttrium iron garnet (YIG) films~\cite{Patton2009,Demo2011,Ando2012,Buch2019}, spin-valve nanocontacts~\cite{Helmut2009} and magnetic vortices~\cite{KSchu2019,KSchu2020}. In all studies, the three magnons are part of the same magnet. Indeed, the magnon interactions are ususally assumed to be very short-ranged. However, this is not self-evident, since the long-range dipolar interaction contributes as well. A nonlocal interaction between different material systems, e.g. a magnon of system A that decays into two magnons in system B, would offer additional functionality for hybrid~\cite{Hoffmann2020} and 3D magnonics~\cite{3D_book}. Here, we demonstrate interlayer three-magnon interactions in a CoFeB$\vert$YIG hybrid nanostructure, in which a magnon in a CoFeB nanowire splits into two counter-propagating magnons in a YIG thin film.

\begin{figure}[ht]
\centering
\includegraphics[width=86mm]{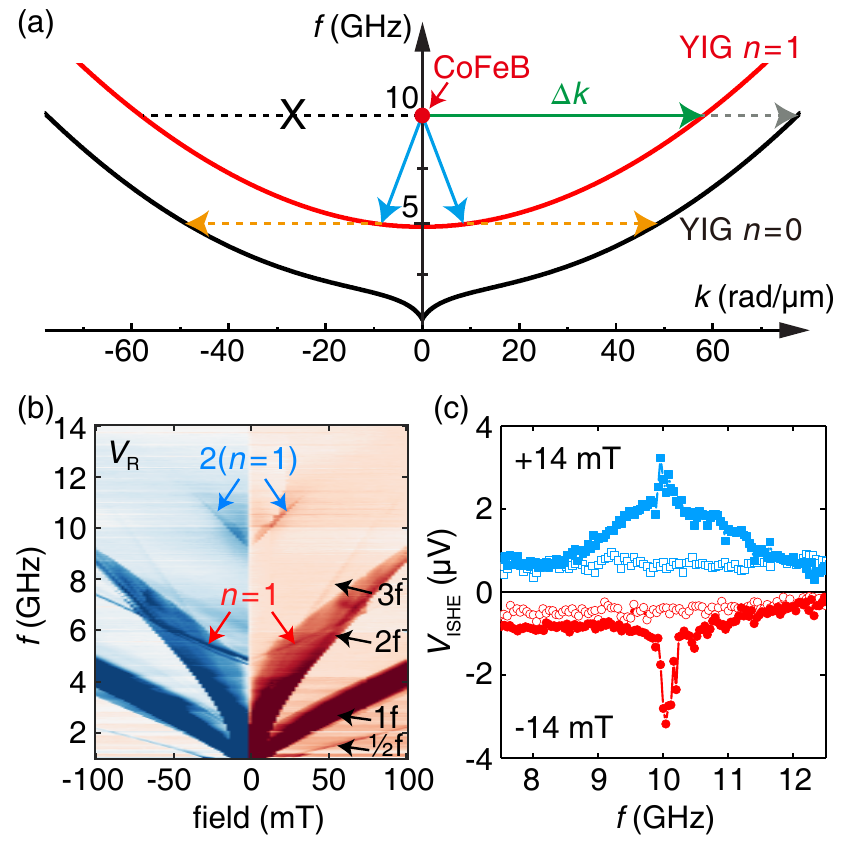} 
\caption{(a) Spin-wave dispersion of the nodeless ($n=0$ black curve) and single-node ($n=1$ red curve) modes for a YIG film with thickness $d=80$~nm and applied field of 10~mT at an angle $\theta=45^{\circ}$. The spin pumping induced by the interlayer two-magnon scattering process (green arrow) is unidirectional, while spin pumping by the interlayer three-magnon scattering (blue arrows) is not. Dashed orange arrows: Interband secondary four-magnon scattering from the $n=1$ to $n=0$ mode. (b) Nonlocal ISHE voltage signals measured at the right Pt electrode $V_{\text{R}}$ at an input power of 32~mW. The black arrows denote excitations of $n=0$ modes at frequencies $\frac{1}{2}f$, $f$, $2f$ and $3f$ modes. The red arrows indicate direct excitation of the $n=1$ modes, while the blue arrows indicate their parametric pumping. The field is swept from negative to positive values. (c) The blue (red) lineplot presents the frequency-dependent ISHE voltage extracted from (b) at an applied field of 14~mT (-14~mT). The blue open squares (red open circles) are extracted under the same conditions for a bare YIG film without CoFeB wire.}
\label{fig2}%
\end{figure}

We excite the magnetic system that consists of a CoFeB nanowire (200~nm wide, 30~nm thick and 100~$\mu$m long) on top of a YIG film with a thickness $d=80$~nm [see Fig.~\ref{fig1}(a)] by the microwaves emitted from a gold stripline antenna~\cite{Ciu2016} (not shown) above. The width of CoFeB wire is characterized by the scanning electron microscope (SEM) shown in the Supplementary Material (SM) Sec.~I~\cite{SI}. The YIG thin films are deposited on gadolinium gallium garnet substrates by radio-frequency magnetron sputtering. We detect propagating magnons by their spin pumping~\cite{Tser2005,Sandweg2011,Madjid2013,Jung2015,Yang2020} into Pt contacts placed on each side of CoFeB at a distance $s=2.5~\mathrm{\mu m}$, in which the inverse spin Hall effect (ISHE) generates a transverse voltage. Figure~\ref{fig1}(b) is a microscopic image of the device. A magnetic field applied at an angle $\theta=45^{\circ}$~\cite{Madjid2013} with respect to the nanowire direction allows both efficient excitation of the nanowire (maximal for 0$^{\circ}$) and detection by ISHE (best for 90$^{\circ}$). Figure~\ref{fig1}(c) shows the ISHE voltage at the left Pt electrode ($V_{\mathrm{L}}$) as a function of excitation frequency and applied field for a small microwave power of 10~mW, which is safely in the linear regime. The red (blue) color represents negative (positive) voltage response. The relatively strong negative ISHE voltage ($\sim-6~\mathrm{\mu V}$, marked by the green arrow) corresponds to the antiparallel magnetization of the CoFeB and YIG layers with large interlayer dipolar coupling~\cite{Chen2018}. Figure~\ref{fig1}(d) shows $V_{\mathrm{L}}$ under high excitation (microwave power 32~mW). We observe additional modes as indicated by the blue and orange arrows in Fig.~\ref{fig1}(d). We argue below that an interlayer three-magnon process causes the former ones and attribute them to parametric pumping of the first excited perpendicular standing spin waves (PSSWs) in YIG by the stray field of the CoFeB Kittel mode [blue arrows in Fig.~\ref{fig2}(a)]. The latter one (orange arrows) should originate from an intralayer four-magnon scattering (orange dashed arrows in Fig.~\ref{fig2}(a)) following the interlayer three-magnon process (blue arrows). The comparison of $V_{\text{L}}$ in Figs.~\ref{fig1}(c) and (d) with $V_{\mathrm{R}}$ in the SM~Sec.~II~\cite{SI} confirms the strong chirality of the linear-response modes (areas marked by green arrows)~\cite{Chen2019,TYu2019}, while the nonlinear signals (blue and orange arrows) are nearly equally strong on both sides. In the vicinity of the CoFeB resonance ($\sim10$ GHz, see microwave reflection spectra $S_{11}$ in the SM~Sec.~III~\cite{SI}), the CoFeB wire switches more easily under high excitation.\\

Figure~\ref{fig2}(a) shows the dispersion of the nodeless ($n=0$) and single-node ($n=1$) perpendicular standing spin waves (PSSWs). The frequency of a spin wave with momentum $k$ in mode $n$ reads~\cite{Slavin1986,Dieterle2019}

\begin{equation}
f_{n}(k)=\frac{\gamma\mu_{0}M_{\mathrm{s}}}{2\pi}\lambda_{\mathrm{ex}}\sqrt{\left[k^{2}+\left(\frac{n\pi}{d}\right)^{2}\right]\left[k^{2}+\left(\frac{n\pi}{d}\right)^{2}+\frac{1}{\lambda_{\mathrm{ex}}}\right]}.
\label{heterosymmetric}%
\end{equation}\\
In calculations, we use the YIG exchange constant $\lambda_{\text{ex}}=3\times10^{-16}\,\mathrm{m}^{2}$, saturation magnetization $M_{\mathrm{s}}=140$~kA/m~\cite{Wu_book} and film thickness $d=80$~nm. According to our modelling explained below, the CoFeB Kittel mode at $\sim$10~GHz couples primarily with the high-$k$ single-node mode as indicated by the green arrow in Fig.~\ref{fig2}(a). This linear interlayer magnon coupling is strongly enhanced in the antiparallel configuration, here in the field interval 0-20~mT~\cite{Chen2018,TYu2019}. The linear process of spin pumping by this \textquotedblleft two-magnon\textquotedblright\ scattering is strongly unidirectional due to the interlayer dipolar interaction~\cite{Chen2019,TYu2019}, with voltage signals in the left Pt electrode $V_{\mathrm{L}}$ [green arrow in Fig.~\ref{fig1}(d)] but not in the right one $V_{\mathrm{R}}$ [Fig.~\ref{fig2}(b)]. In this process, the nanometric width of CoFeB wire generates a broad $k$ distribution and thus enables efficient scattering between a $k=0$ CoFeB magnon and a high-$k$ $n=1$ YIG magnon given by the dispersion in Fig.~\ref{fig2}(a). We derive the interlayer magnon-magnon coupling strength by the magnetodipolar interaction for the $n=0$ and $n=1$ modes with $+k$ and $-k$ wave vectors in the SM~Sec.~IV-C~\cite{SI}. The chiral spin pumping signal scales linearly with the microwave power and can be detected down to 1~mW (see SM~Sec.~V~\cite{SI}). At powers above 10~mW, nonlinear effects emerge. Figure~\ref{fig2}(b) shows the spin pumping signals measured at the right Pt electrode at 32~mW. We observe multiple new features associated with the $n=0$ mode including parametric pumping of the $f$ mode at $2f$ microwave excitation~\cite{Sandweg2011} and a triple-frequency ($3f$)~\cite{Helmut2009,Wol2022} mode, but also second harmonic generation at microwave frequencies $\frac{1}{2}f$. The $n=1$ PSSW mode is observed in Fig.~\ref{fig2}(b) indicated by the red arrows. By varying the film thickness from 80~nm to 40~nm, the $n=1$ mode shifts to $\sim$12~GHz as shown in the SM~Sec.~VI~\cite{SI}. Evidence for parametric pumping of the higher $n=1$ mode appears at high frequencies around 10 GHz as marked by blue arrows. Figure~\ref{fig2}(c) shows two lineplots at +14~mT (blue squares) and -14~mT (red dots) with positive and negative ISHE voltage signals around 10~GHz. Open blue squares and open red circles show data obtained from a bare YIG sample without CoFeB wire on top (see SM~Sec.~VII~\cite{SI}). If we replace the 200~nm-wide CoFeB wire by a 800~nm-wide one, the CoFeB Kittel mode frequency drops significantly and no longer matches twice the frequency of the $n=1$ mode. As a result, no signal is observed around 10~GHz (see SM~Sec.~VII~\cite{SI}). The signals in Fig.~\ref{fig2}(b) marked by blue arrows are therefore caused by parametric pumping of $n=1$ YIG magnons by the stray field from the CoFeB dynamics but not the stripline.\\

Our experiments uniquely combine the advantages of microwave and electrical magnon transport studies. The observable is $S_{21}\left(\omega_{1},\omega_{2}\right)$, the bichromatic scattering matrix of a magnon injected at frequency $\omega_{1}$ at contact/stripline $1$ to a magnon with frequency $\omega_{2}$ at contact/stripline $2$. Propagating magnon spectroscopy~\cite{Vla2008} studies the coherent magnons at frequency $\omega,$ in terms of $S_{21}\left(\omega,\omega\right)$. The electrical injection and detection of magnons by heavy metal contacts~\cite{vanWees2015} is the method of choice to study diffuse magnon transport. However, senses only $\int\left\vert S_{21}\left(\omega_{1},\omega_{2}\right)\right\vert d\omega_{1}d\omega_{2},$ so all spectral information is lost. Here we measure the coherent response to inductive magnon injection at frequency $\omega$ and electric
detection at a distant contact, i.e. $\int\left\vert S_{21}\left(\omega,\omega_{2}\right)\right\vert d\omega_{2}$. In the linear regime this does not provide new information. However, the emergence of a magnon frequency comb leads to an increased signal at a magnon resonance $\omega=\omega_{k}$, while new signals due to parametric pumping emerge when $\omega=2\omega_{k}$. When the magnon decay length is larger than the contact distance, we can interpret the experiments simply in terms of the magnon spectrum generated by the microwaves under the stripline since the electrical detection is not sensitive to the propagation phase. Here we model the observed non-linearities by the leading terms in the Holstein-Primakoff expansion with Hamiltonian

\begin{equation}
\mathcal{\hat{H}}=\mathcal{\hat{H}}_{\mathrm{C}}^{\left(  0\right)  }+\mathcal{\hat{H}%
}_{\mathrm{Y}}^{\left(  0\right)  }+\mathcal{\hat{H}}_{\text{\textrm{CY}}}%
^{\mathrm{3M}},\label{Hamiltonian}%
\end{equation}\\
Here $\mathcal{\hat{H}}_{\mathrm{C}}^{\left(0\right)}=\varepsilon_{\mathrm{C}}c^{\dag}c$ and $\mathcal{\hat{H}}_{\mathrm{Y}}^{\left(  0\right)  }=\sum_{\mathbf{k}n}\varepsilon _{\mathbf{k}n}a_{\mathbf{k}n}^{\dag}a_{\mathbf{k}n}$ are the excitations of the Kittel mode in the magnetic wire and spin waves in mode $\mathbf{k}n$ of the film. In principle, all states may be excited by the microwaves emitted by the stripline with mode-dependent efficiencies. The leading non-linear term is the 3-magnon interaction $\mathcal{\hat{H}}_{\text{\textrm{CY}}}^{\mathrm{3M}}$. In the following, we model the nonlinear excitations observed around the CoFeB nanowire resonance frequency (10~GHz) by the magneto-dipolar field of the nanowire in the YIG thin film with an interlayer 3-magnon interaction $\mathcal{\hat{H}}_{\text{\textrm{CY}}}^{\left\langle\mathrm{3M}\right\rangle}=\mathcal{D}_{\vec{k}^{+}\vec{k}^{-}}^{(n)}\hat{c}^{\dag}\hat{a}_{n,\vec{k}^{+}}\hat{a}_{n,\vec{k}^{-}}+\mathrm{H.c}.$, where $\hat{c}$~($\hat{a}_{n,\vec{k}^{\pm}}$) is the annihilation operator of the CoFeB nanowire Kittel mode magnon (YIG magnon of $n=0,1$ with wave vector $\vec{k}^{\pm}$), $\mathcal{D}_{\vec{k}^{+}\vec{k}^{-}}^{(n)}$ is a coefficient, and $\vec{k}^{+}\approx-\vec{k}^{-}$ (see SM~Sec.~IV-A~\cite{SI}). The efficiency of the parametric excitation scales with the ellipticity of the excited magnon pairs, which decreases with $k$, so $|\mathcal{D}_{\vec{k}^{+}\vec{k}^{-}}^{(1)}|>|\mathcal{D}_{\vec{k}^{+}\vec{k}^{-}}^{(0)}|$. Therefore, the CoFeB Kittel mode excites $n=1$ YIG magnon pairs at a lower threshold than that of $n=0$ pairs. The parallel magnetic pumping by the Zeeman interaction $\mu_{0}\gamma\vec{m}_{\Vert}^{\mathrm{YIG}}\cdot\vec{h}_{\mathrm{dip}}$ does not depend on the polarization of neither the magnon nor the dipolar field, in contrast to the chiral spin pumping~\cite{Chen2019,TYu2019}. The 3-magnon interaction is therefore not chiral and the signals in both Pt contacts are nearly the same [see Figs.~\ref{fig2}(b) and (c)].

\begin{figure}[ht]
\includegraphics[width=86mm]{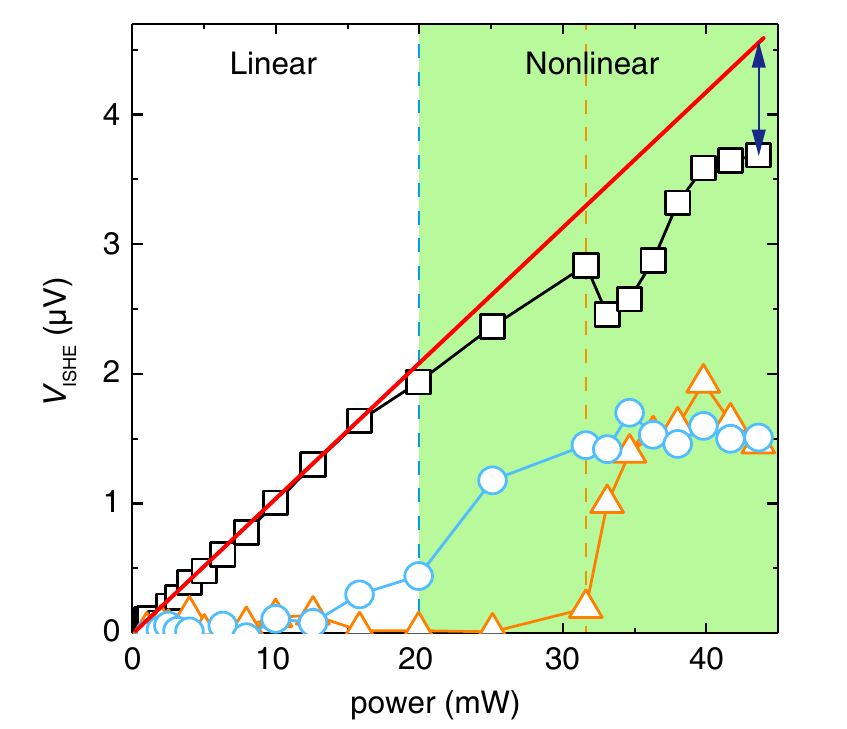}  
\caption{ISHE peak voltages of the right Pt electrode as a function of the input microwave power measured at
5~GHz for the low-$k$ $n=1$ PSSW mode (black open squares) and at 10~GHz for the nonlinear modes (light blue open circles and orange open triangles). The red line is a linear fit up to 20~mW. The dark blue double-sided arrow marks the deviation from linearity. The light green area represents the nonlinear regime of interlayer three-magnon processes (light blue open circles) above 20~mW (blue dashed line). An additional feature (orange arrows in Fig.~\ref{fig1}(c) and orange open triangles) observed above 32~mW (orange dashed line) is attributed to a secondary four-magnon process as indicated by orange dashed arrows in Fig.~\ref{fig2}(a).}
\label{fig3}
\end{figure}
In Fig.~\ref{fig3} we address the power dependence of the signals at 5~GHz and 10~GHz. We focus on the $V_{\mathrm{ISHE}}$ on the right Pt electrode for input powers from 1~mW to 47~mW (see SM~Sec.~VIII~\cite{SI} for the raw data). The signal attributed to the interlayer three-magnon interaction at microwave powers above 20~mW (light green area in Fig.~\ref{fig3}) are nearly the same in both contacts. The signal associated to the excitation of the low-$k$ $n=1$ mode deviates from a linear power dependence (black open squares) at $\sim$20~mW, which we interpret an evidence for the Suhl instability~\cite{Suhl1957}. The spin pumping signal drops above the threshold, because the confluence scattering opposes the Kittel magnon decay in the wire. When the power reaches 32~mW, an additional mode (orange arrows in Fig.~\ref{fig1}(d)) emerges. We attribute this additional mode to a four-magnon process~\cite{Helmut2012,Moh2021} during which two $n=1$ YIG magnons scatter to two $n=0$ YIG magnons, i.e. $a_{1}^{\dag}a_{1}^{\dag}a_{0}a_{0}+\mathrm{H.c}.$ (orange dashed arrows in Fig.~\ref{fig2}(a)). The drop in the $n=1$ mode intensity (black open squares) accompanies a new signal (orange open triangles), similar to a four-magnon scattering signal reported for a YIG nanoconduit~\cite{Moh2021}. We explain the increased slope of the orange mode as a function of field as follows, see also SM~Sec.~IV-B~\cite{SI}. The four-magnon interaction scales like $\propto|\vec{k}^{\prime\pm}|^{2}$, where $\vec{k}^{\prime\pm}$ are the momenta of the $n=0$ magnons degenerate with the $n=1$ magnons that are efficiently excited by the CoFeB Kittel mode when their $\vec{k}^{\pm}$ is small. The amplitude of the $n=1$ magnons therefore decreases with excitation frequency higher than $\approx2\min\omega_{1,\vec{k}}$, but $|\vec{k}^{\prime\pm}|$ of the $n=0$ magnons increases. The secondary maximum of the spin pumping signals seen in experiments and calculations reveals that the four-magnon scattering can win this competition in a narrow frequency interval. We do not observe indications for an intralayer three-magnon process in which one $n=1$ YIG magnon splits into two $n=0$ YIG magnons, because the overlap integrals are suppressed due to the different parity of the standing wave amplitudes (see SM~Sec.~IV-B~\cite{SI}). Finally, we demonstrate in Fig.~S6 of the SM~\cite{SI} excellent agreement of the calculated resonance energies with the observed spectra at both low and high excitation powers.\\

In conclusion, we detect nonlinear interlayer magnon interactions in a hybrid magnetic nanostructure (YIG$\vert$CoFeB) by nonlocal spin pumping. The leading nonlinearity is a three-magnon process in which one CoFeB Kittel magnon splits into a pair of single-node ($n=1$) YIG magnons with opposite wave vectors ($+k$ and $-k$). By comparing the ISHE voltage signals of left and right Pt electrodes, we find nearly symmetric magnon emission in both directions in contrast with the almost perfect chirality of linear excitations in agreement with model calculations based on purely magnetodipolar couplings. The theoretical analysis also
indicates that the nonlinear interlayer coupling with single-node ($n=1$) YIG magnons dominates over that with nodeless ($n=0$) ones. We attribute an additional signal at even higher power to a cascade of \textit{interlayer} three-magnon and \textit{intralayer} four-magnon processes. Understanding the dynamics in hybrid magnetic systems may help engineer dissipation and cross talk in nanomagnonic devices, which is a necessary step in the prospect of quantum magnonics for entanglement distillation through nonlinear coupling of local nanowire magnons and paires of long-distance propagating magnons.\\

\section{ACKNOWLEDGMENTS}

We thank D. Wei for helpful discussions. The authors acknowledge support from the NSF China under Grants 12074026, 12104208 and U1801661, the National Key Research and Development Program of China Grants 2016YFA0300802 and 2017YFA0206200, and JSPS Kakenhi Grants \# JP19H00645 and 21K13847.

\providecommand{\noopsort}[1]{}\providecommand{\singleletter}[1]{#1}

\end{document}